\documentclass[journal]{IEEEtran}
\usepackage {graphicx}
\usepackage{subfigure}
\DeclareGraphicsExtensions{.eps,.pdf,.jpeg,.png}
\usepackage[cmex10]{amsmath}
\usepackage{amssymb}
\usepackage{cite}
\usepackage{multirow}
\usepackage{array} 
\begin{document}

\title{On the Capacity Region of Multiple Access Channel }

\author{Xuezhi Yang,~\IEEEmembership{Senior Member,~IEEE }

\thanks{Xuezhi Yang is the inventor of several 4G fundamental techniques including  soft frequency reuse, scalable OFDM and random beam forming. His new invention, multi-level soft frequency reuse,  can improve the overall spectrum efficiency of 4G network by 30\% with near zero cost. He is  seeking the chance of cooperation with the industry. He is located in  Beijing, China  (email: yangxuezhi@hotmail.com) }
\thanks{}}

\markboth{}{}

\maketitle


\begin{abstract}

The capacity region of a multiple access channel is discussed. It was found that orthogonal multiple access and non orthogonal multiple access have the same capacity region under the constraint of same sum power.

\end{abstract}



\section{Introduction}

Multiple access techniques are used to allow a  number of mobile users to share the same spectrum.  FDMA, TDMA, CDMA and OFDMA are used in the 1th to 4th generation mobile communication systems, all in an orthogonal way.  With the coming of 5G, non orthogonal multiple access (NOMA) became popular, both in the downlink \cite{DownlinkNOMA} and uplink\cite{UplinkNOMA}. 

The capacity region of multiple access channel (MAC) or uplink channel  is well known \cite{Andrea,DavidTse}. The two user model of  MAC is described as

\begin{equation}
y=x_1+x_2+n,
\end{equation}
where $x_1$ and $x_2$ are signals of user one and two, with the power of $P_1$ and $P_2$, $n$ is the white noise with power $N$, $y$ is  the received signal.  

\begin{figure}[!ht]
\begin{center}
\includegraphics[width=3in]{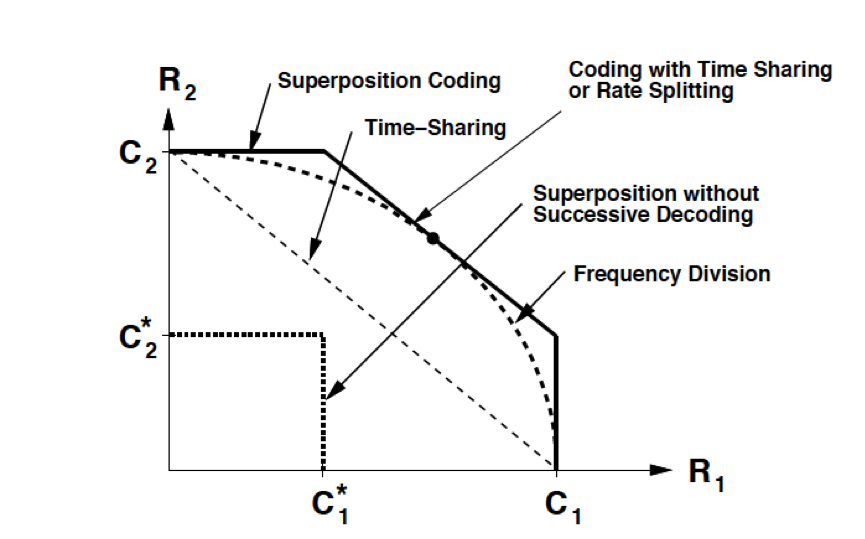}
\caption{Two-User MAC Capacity Region\cite{Andrea}.}
\label{Fig-PTx2}
\end{center}
\end{figure}

Let $R_1$ and $R_2$ be the reachable data rate of user one and two, the capacity region of such a model is represented by

\begin{equation}
R_1\leqslant \log(1+\frac{P_1}{N})=C_1,
\end{equation}

\begin{equation}
R_2\leqslant \log(1+\frac{P_2}{N})=C_2,
\end{equation}

and

\begin{equation}
R_1+R2\leqslant \log(1+\frac{P_1+P_2}{N}).
\end{equation}

This region is illustrated in Fig.  \ref{Fig-PTx2} by the solid line. Notice that  $x_1$ and $x_2$ share the same resource,  such a way of communication is called superposition coding. 

The decoding of superposition coding relies on a technique called successive interference cancellation (SIC).  Let's take the rate point $(C_1, C_2^*)$ in  Fig.  \ref{Fig-PTx2} as an example. In the encoding phase, we can put $x_1$ with power $P_1$ on the channel first, yielding a rate $C_1$, then put $x_2$ with power $P_2$ on the channel, treating $x_1$ as noise, yielding a rate $C_2^*$,

\begin{equation}
C_2^*= \log(1+\frac{P_2}{P_1+N}).
\end{equation}

In the decoding phase, $x_2$ is decoded first and then reconstructed and eliminated from $y$,  so $x_1$ can be decoded successfully without the interference of $x_2$. 

If the two users are time divided (TD) , suppose user one is allocated a fraction  $\alpha\in [0,1]$ of the whole time, and the rest of time is allocated to user two, then the sum rate

 \begin{equation}
R_1+R2 \leqslant \alpha C_1+ (1-\alpha)C_2.
\end{equation}

If two users are frequency divided (FD) and $B$ is the total bandwidth,  user one is allocated $\alpha B$ and $(1-\alpha)B$ is assigned to user two, the capacity region is

 \begin{equation}
R_1+R2 \leqslant  \alpha \log (1+\frac{P_1}{\alpha N})+ (1-\alpha)\log(1+\frac{P_2}{(1-\alpha)N}),
\end{equation}
which is illustrated by the dashed curve in Fig.  \ref{Fig-PTx2}. Notice there is one point  FD can reach the same sum rate as superposition coding. On this point, the power of each user is proportional to the bandwidth allocated. 

\section{Discussions}

Based on the former information, it was widely believed that superposition coding is the way to achieve maximum capacity, dominating orthogonal measures as TD or FD. However, this superiority exists only under the former stated constraints, i.e. the power of the two users are $P_1$ and $P_2$. This constraint is not reasonable in practical cases, since a mobile station usually does not transmit  at full power and can adjust its transmit power according to its position and data rate. 

Then, if we relax the constraint of certain power for each user to certain sum power for two users, the capacity region of superposition coding becomes
 
\begin{equation}
R_1+R2\leqslant \log(1+\frac{P_1+P_2}{N}).
\end{equation}

For the TD case, in any time slot,  one user use the power $P_1+P_2$ and the other keep silent, the sum power is $P_1+P_2$. For the FD case, if user one is allocated $\alpha B$ with power $\alpha(P_1+P_2)$, and user two is allocated  $(1-\alpha)B$ with power $(1-\alpha)(P_1+P_2)$, the sum power constraint is also satisfied. It can be easily verified that TD and FD have the same capacity region as superposition coding in MAC.

\section{Conclusion}

Under the constraint of sum power, TD, FD and superposition coding have the same capacity region in MAC. So NOMA is not an option in the 5G uplink for the capacity reason.

\bibliographystyle{IEEEtran}
\bibliography{IEEEfull,MAC}

\begin{thebibliography}{1}
\providecommand{\url}[1]{#1}
\csname url@samestyle\endcsname
\providecommand{\newblock}{\relax}
\providecommand{\bibinfo}[2]{#2}
\providecommand{\BIBentrySTDinterwordspacing}{\spaceskip=0pt\relax}
\providecommand{\BIBentryALTinterwordstretchfactor}{4}
\providecommand{\BIBentryALTinterwordspacing}{\spaceskip=\fontdimen2\font plus
\BIBentryALTinterwordstretchfactor\fontdimen3\font minus
  \fontdimen4\font\relax}
\providecommand{\BIBforeignlanguage}[2]{{%
\expandafter\ifx\csname l@#1\endcsname\relax
\typeout{** WARNING: IEEEtran.bst: No hyphenation pattern has been}%
\typeout{** loaded for the language `#1'. Using the pattern for}%
\typeout{** the default language instead.}%
\else
\language=\csname l@#1\endcsname
\fi
#2}}
\providecommand{\BIBdecl}{\relax}
\BIBdecl

\bibitem{DownlinkNOMA}
Y.~Saito, Y.~Kishiyama, A.~Benjebbour, T.~Nakamura, A.~Li, and K.~Higuchi,
  ``Non-orthogonal multiple access (noma) for cellular future radio access,''
  in \emph{Proc. Vehicular Technology Conference (VTC Spring), 2013 IEEE 77th},
  June 2013, pp. 1--5.

\bibitem{UplinkNOMA}
M.~Al-Imari, P.~Xiao, M.~A. Imran, and R.~Tafazolli, ``Uplink non-orthogonal
  multiple access for 5g wireless networks,'' in \emph{Proc. 2014 11th
  International Symposium on Wireless Communications Systems}, Aug. 2014, pp.
  781--785.

\bibitem{Andrea}
A.~Goldsmith, \emph{WIRELESS COMMUNICATIONS}.\hskip 1em plus 0.5em minus
  0.4em\relax Cambridge Univ. Press, 2005.

\bibitem{DavidTse}
D.~Tse and P.~Viswanath, \emph{Fundamentals of Wireless Communication}.\hskip
  1em plus 0.5em minus 0.4em\relax Cambridge Univ. Press, 2005.

\end{thebibliography}

\end{document}